# Enhanced Phase Sensitive SD-OCT for flow imaging using ultrasonically sculpted optical waveguides

LLOYD LOBO[1], JUNZE LIU[2], HANG YANG[2], YASIN KARIMI[1], B. HYLE PARK[2], AND MAYSAMREZA CHAMANZAR[1,*]

[1]*Electrical and Computer Engineering Department, Carnegie Mellon University, Pittsburgh, PA 15213, USA*
[2]*Department of Bioengineering, University of California, Riverside, CA 92521, USA*
*\*mchamanz@andrew.cmu.edu*

**Abstract:** Phase sensitive detection in spectral domain optical coherence tomography (SD-OCT) is a powerful method for functional imaging of biological events with high spatiotemporal resolution. The depth-dependent signal-to-noise ratio (SNR) is a limiting factor on the minimum detectable phase changes of phase in shot noise-limited SD-OCT systems. The SNR over a depth is constrained by the terminal optics, usually using a focusing lens to project light into the tissue and collect the backscattered light. In situ ultrasonically sculpted optical waveguides have been used to improve SNR roll-off over depth compared to conventional SD-OCT systems. In this paper, we extend this feature to demonstrate phase sensitive detection at depth using ultrasonically enhanced OCT (ue-OCT). Our experimental results show that ultrasonically sculpted optical waveguides are phase stable and follow near shot-noise limited behavior. We measured milk flow velocity changes to demonstrate a phase sensitivity of 5.25 mrad at 10 dB SNR and dynamic range of 0.8 mm/s to 14.7 cm/s using ue-OCT. Our results show flow detection with ue-OCT at extended depths (i.e., 3.5 mm) otherwise not possible with conventional SD-OCT systems with matched focal lengths. The results in this paper show the potential of ue-OCT for phase-sensitive flow measurement from the depth of tissue for a gamut of applications such as cerebral blood flow imaging as a proxy to neural activity mapping.

## 1. Introduction

Optical imaging technologies are well-suited to image biological tissue. Optical techniques employ variation in reflectance, absorption and scattering properties within biological samples to obtain structural information of the samples non-invasively with high resolution. For increased contrast, interferometric optical techniques like optical coherence tomography (OCT) have been developed to better visualize internal structures of tissue[1,2]. Spectral-Domain OCT (SD-OCT) implementations have augmented the capabilities of time domain OCT systems by further increasing the sensitivity to tissue optical properties[3–7]. In addition, SD-OCT spectrally samples the tissue depth and has enabled functional imaging of biological processes. The advancements in SD-OCT allow for *in situ* visualization of tissue samples, with ~10 μm-scale resolution, for oxygen saturation levels in blood, blood flow rate changes and neural activity [8–12]. Dynamic changes in the tissue such as blood flow cause doppler frequency changes that can be detected using phase-sensitive SD-OCT[13,14].

While SD-OCT is a powerful technique for detecting functional activity, the imaging depth in tissue is limited due to attenuation of light intensity through the depth. The attenuation of light intensity through tissue is because of absorption, scattering and geometrical expansion of the light beam through the tissue. These effects have limited SD-OCT to visualizing blood flow changes only at shallow depths within tissue. To mitigate attenuation due to light scattering, longer wavelengths of light 1.3 - 1.7 μm have been used in SD-OCT systems to enable detection of flow rate changes at a depth of 1.4 mm [15–17]. However, this comes at the cost of increased absorption through the tissue due to the water content in the tissue and exacerbated geometrical effects, such as reduced axial and lateral resolution[18,19]. Conventional SD-OCT imaging systems use external Gaussian lenses to project and collect light through the tissue. To reduce the geometrical expansion of light through the tissue, the Rayleigh range of the externally focused optical beam must be increased. There is a fundamental trade-off between the lateral optical spot size and the Rayleigh range along the axial direction for a Gaussian lens[20]. This trade-off limits the signal to noise ratio (SNR) of the retrieved light. Either the peak SNR from a particular depth of interest can be optimized (small spot size, but limited Rayleigh range) or the depth range over which high SNR can be obtained is optimized, at the expense of an overall larger spot size and lower SNR. Axicons can be used to extend the Rayleigh range by generating Bessel-shaped optical beams[21,22] as opposed to Gaussian beams. However, due to the inefficient distribution of optical energy into the central spot, the system sensitivity is significantly reduced (~18 dB). Adaptive optics techniques have been used to augment the sample arm of SD-OCT systems to extend high SNR regions by utilizing tunable lenses for multi-focal plane selectivity [23–26]. While these techniques leverage the tunability of electro-optic and acoustic-optic modulators to achieve multi-focal plane selectivity, there exists an inherent tradeoff between focal depth and lateral resolution in these systems. In addition to this, these techniques often compromise imaging speed. Alternatively, synthetic aperture techniques also offer extension in the Rayleigh range but remain difficult to implement experimentally and drastically increase system complexity and cost[27–29].

We have previously demonstrated ultrasonically-enhanced OCT (ue-OCT) by integrating *in situ* ultrasonically sculpted gradient refractive index (GRIN) optical waveguides into an SD-OCT system to extend the Rayleigh range without compromising the lateral resolution (Fig. 1) [30,31]. The *in situ* ultrasonically sculpted GRIN optical waveguides and guide light through a sample and enables *in situ* lensing. The advantage of ue-OCT over conventional lensing in SD-OCT augers enhancements for structural imaging[30]. In this work, we further expand on these results to show that ue-OCT can also enable phase sensitive detection through an extended depth of the sample. We characterize the phase stability of our ue-OCT system to showcase viability of such systems for functional imaging via phase-sensitive SD-OCT. We exemplify this by imaging milk flow, as an optically scattering liquid[32,33]. We demonstrate that milk flow rate changes similar to blood flow changes, corresponding to functional neural activity in the brain, can be detected by our ue-OCT system. We also demonstrate improved flow detection in ue-OCT when compared with conventional SD-OCT through an extended sample depth.

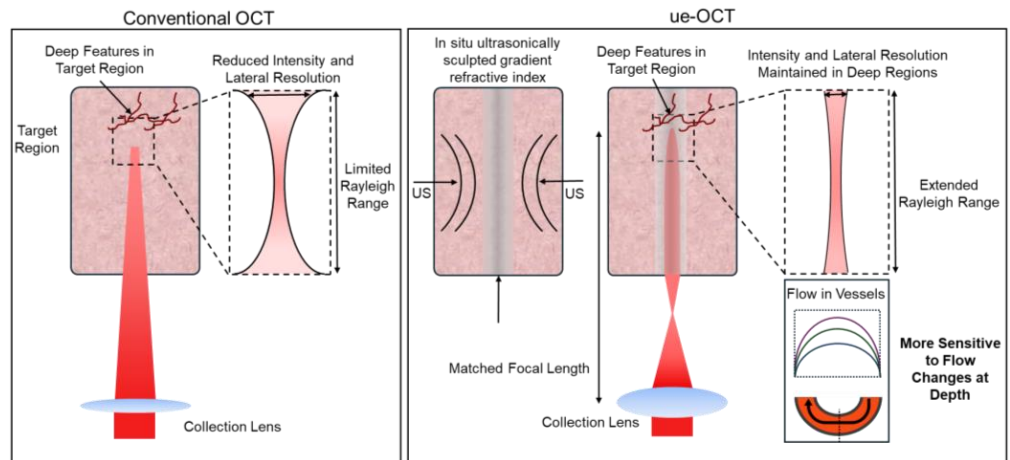

*Figure 1:* Gradient refractive index fields can be sculpted inside a sample using ultrasound waves. Integrated within the sample arm in OCT, we develop ue-OCT. ue-OCT provides extended Rayliegh Range compared to conventional OCT and is more sensitive to flow changes at depth

## 2. Ultrasonically Enhanced OCT System Design

We designed a conventional SD-OCT system and augmented it with an in situ ultrasonically sculpted optical lens to realize a ue-OCT system. We rigorously compared the SNR roll-off, phase stability and flow detection between these systems. Our ue-OCT system, illustrated in Fig. 2.a), was implemented with a spectrometer, and a superluminescent diode (SLD) with a center wavelength $\lambda_0$ of 830 nm and optical power of 20 mW (SLD830S-A20W; Thorlabs Inc. USA) as the optical source. The SLD is housed in a commercial current driver (CCS-std; Aerodiode Co., France) to modulate light in synchronization with the beam-shaping ultrasound transducer. The source has a 3dB optical bandwidth of 55 nm. The source is fed to a 50:50 fiber coupler (TW850R5A2; Thorlabs Inc. USA) through an isolator (IO-F-850APC; Thorlabs Inc. USA) to prevent damage from back-reflected power. The fiber coupler routes light to the reference and sample arms of the SD-OCT system. Light collected by these two arms is coupled to the spectrometer arm for detection. The spectrometer contains a fiber collimator (C80APC-B; Thorlabs Inc. USA) that directs light towards a diffraction grating (1800 l/mm @ 840 nm; Wasatch Photonics Inc., USA). for spectral-spatial mapping of different wavelengths of light. The wavelength-separated beam is illuminated onto a line-scan CMOS camera (spL4096-140 km, 4096 pixels; Basler AG Co., Germany) via a focusing lens (AC508-100-AB; Thorlabs Inc. USA). The reference arm of the system contains a polarization controller to match the polarization of the reflected light from the sample arm for maximum interference between the two arms. The reference and sample arms are designed in free space to allow for interchangeable focusing optics combined with various sample targets. The reference arm optical beam is incident on a mirror through a variable neutral density filter (NDC-50C-2; Thorlabs Inc. USA) through a fiber port (PAF-X-11-B; Thorlabs Inc. USA). A glass tank filled with water is used to allow for efficient ultrasound coupling into the target sample. The reference arm mirror is placed within the water to reduce dispersion-related deviation between the sample and reference arms.

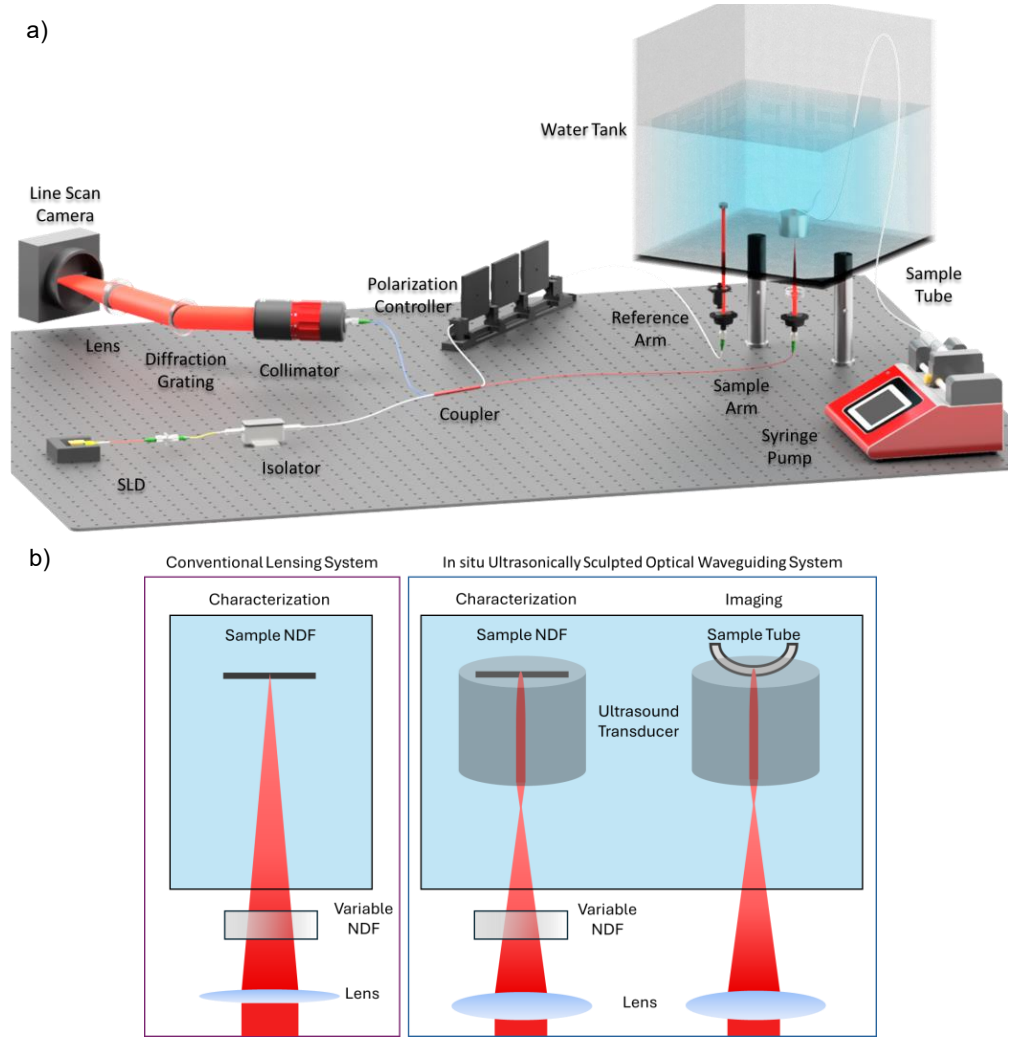

*Figure 2:* a) 3D-Schematic of ultrasonically enhanced SD-OCT setup; b) Sample arm configurations for system characterization and flow imaging

The ue-OCT system deviates from conventional systems in the sample arm. In conventional SD-OCT systems, a single achromatic focusing lens ($f_l$ = 60 mm) is used to deliver and collect light to and from the target sample (Fig. 2.b). To implement ue-OCT, we use a cascade of a focusing lens ($f_l$ = 30 mm) and an *in situ* ultrasonically sculpted GRIN optical waveguide to deliver and collect light to and from the target sample. The sample arm focal length in both the SD-OCT and ue-OCT systems are matched for fair and quantifiable comparisons of Rayleigh range and spot size. The ultrasound transducer (Physik Instrumente) that creates the ultrasonic waveguide has an outer diameter of 40 mm, a length of 30 mm, and a wall thickness of 1 mm. The nominal fundamental thickness mode resonant frequency of the transducer is 2.125 MHz. The *in situ* ultrasonically sculpted GRIN optical waveguide is formed by a time-averaged standing wave generated by the ultrasonic transducer. The optical source must be synchronized with the positive crest of the standing wave[34]. The voltage applied across the ultrasound transducer is selected to form a focal plane at the desired target depth within the sample. A 2-channel function generator with an internal trigger (SDG6022X, Siglent Technologies) is used to synchronize the light source and the positive refractive index gradient sculpted along the optical path.

The duty cycle of the optical source determines the effective waveguiding within one A-line acquisition. The ultrasonically sculpted GRIN optical waveguide is formed by the average refractive index gradient during the 'ON' period of the optical source. Therefore, by changing the duty cycle, the resultant waveguiding is changed. The duty cycle used in our ue-OCT implementation is selected by the maximum SNR at a single depth (Fig. 3). A reflective ND filter is used as the sample. The duty cycle and synchronized is swept to find the optimal operating condition. While the duty cycle is increased, the increased photon count is counteracted by the smearing of focus along depth. The resultant SNR obtained from the OCT interferogram is maximized when the competing effects of increased photon count and effective waveguiding are optimized (Fig. 3). The ue-OCT system is operated with the optical source at 6% duty cycle as the SNR is maximized.

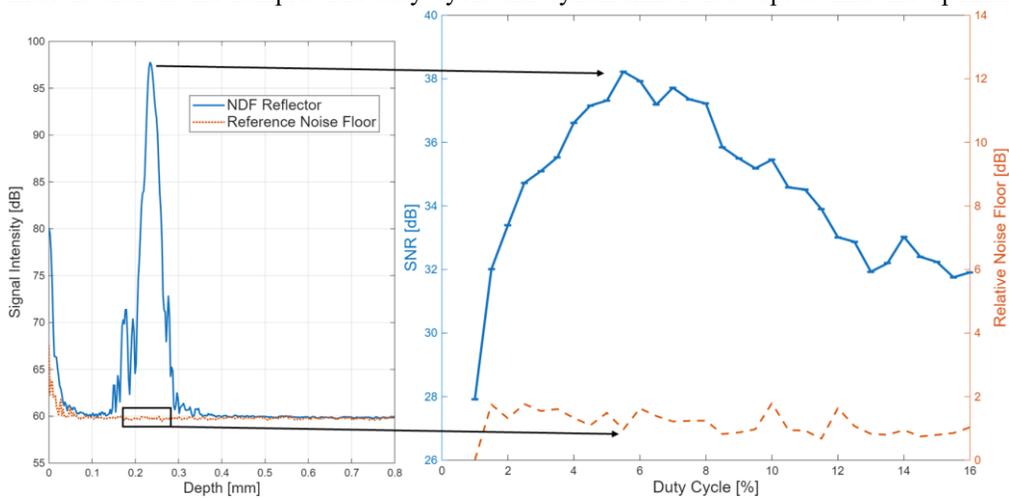

*Figure 3:* A 6% duty cycle results in maximum SNR for *in situ* ultrasonic GRIN optical waveguiding in ue-OCT

## 3. Increased Rayleigh range for Enhanced SNR roll-off

The SNR roll-off in a spectrometer-based SD-OCT system is determined by several different factors, including the spectrometer design and the focusing optics used to acquire a depth profile (A-line). Our work addresses the limitations imposed by external focusing optics in the sample arm of an SD-OCT system. Conventional lenses have a fixed focal length ($f$), implying that for a fixed beam diameter ($D$), the Rayleigh range ($z_R$) is fixed as described by Eq. 1[20]:

$$z_R = \frac{4\lambda_0 f^2}{\pi n D^2}, \quad (1)$$

where $n$ is the refractive index of the medium and $\lambda_0$ is the center wavelength of the optical source. The Rayleigh range is defined as the distance from the minimum beam waist where the beam waist increases by a factor of $\sqrt{2}$. This is desired to be as long as possible for OCT applications. However, the tradeoff between the beam spot size and the Rayleigh range in conventional optical lenses, renders the Rayleigh range limited.

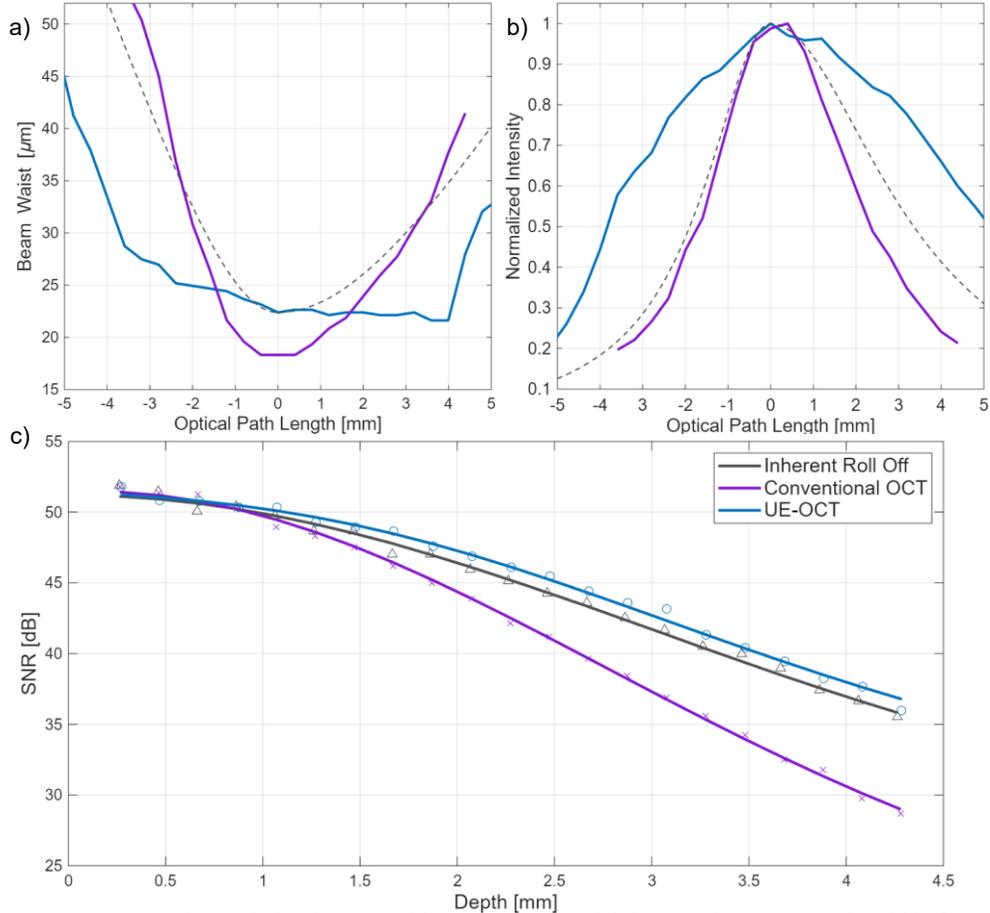

Figure 4: a) *in situ* ultrasonic lensing can achieve a greater Rayleigh range than a conventional ideal diffraction-limited lens; b) The normalized forward path intensity is maintained through depth using *in situ* ultrasonic lensing; c) The focusing optics in ue-OCT exhibits a slower OCT SNR roll off compared to conventional ideal diffraction-limited lenses with the same effective focal length.

In our ue-OCT implementation, we use an *in situ* ultrasonically sculpted gradient refractive index (GRIN) optical waveguide to guide light and induced lensing. ue-OCT alleviates this tradeoff and enables an extended Rayleigh range without significantly compromising the beam spot size (Fig. 4.a-b). The confocal parameter ($2z_R$) is extended from 3.9 mm to 8.5 mm. It can be observed that the beam waist created by the conventional lens is slightly smaller than that formed by the *in situ* ultrasonically sculpted GRIN optical waveguide. A theoretically matched beam waist with a Gaussian beam profile (Fig. 4.a-b) shows the improvement is not only due to the marginally larger beam waist formed by the ultrasonically sculpted optical waveguide, but rather due to the Bessel-shaped GRIN profile integrated in ue-OCT. The measured beam waists display a lopsided trend around the minimum beam waist because of refractive index changes associated with different depth regions along the sample.

Extension of the Rayleigh range is manifested in the depth dependent SNR roll-off in the OCT depth intensity profiles. To characterize the SNR along depth for each system, we used a reflective neutral density (ND) filter (ND520B; Thorlabs Inc. USA) which provides a strong reflection off its coated surface. We show in Fig. 4.c) the extension of the Rayleigh range realized as an improvement in SNR roll-off between the two systems. The ND filter is initially placed at the focal plane of the focusing lens ($f_2$ = 60 mm). The reference arm length is then stepped by 0.2 mm using a linear translation stage (LTS150; Thorlabs Inc. USA) to obtain the spectrometer-limited SNR roll-off. In the case of the cascade system, the voltage that is supplied to the ultrasonic transducer is tuned to maximize the back-reflected signal from the sample at the initial depth position. The sample is then scanned over an optical path length of 4 mm with 0.2 mm steps using a linear translation stage (LTS150; Thorlabs Inc. USA). The reflection from the sample ND filter is compared at different depths to characterize the SNR roll-off along depth for each focusing optic configuration. The focal lengths for both systems are matched, and the initial SNR is adjusted for a fair comparison. The extracted SNR values are initially fit to a spectrometer related depth-dependent SNR roll off curve followed by an additional Gaussian term to account for the SNR roll-off due to lensing in the sample arm [35]. The intensity of the peak at each depth is captured by 1024 A-scans and is averaged to determine the nominal SNR. The ue-OCT system offers a 7.8 dB SNR improvement over a 4 mm depth range compared with a conventional SD-OCT.

## 4. ue-OCT is as Phase Stable as Conventional SD-OCT

The phase noise of an OCT system determines the minimum detectable phase change [36,37]. Phase noise in shot-noise-limited OCT systems can be expressed by Eq. 2:

$$\sigma_{\Delta\phi} \propto \frac{1}{\sqrt{SNR}} . \qquad (2)$$

Apart from shot-noise, other noise sources like bulk system mechanical vibrations and relative intensity noise (RIN) raise the minimum detectable phase change at a given SNR. Common-mode noise sources that are independent of the targeted phase changes are dampened by using a differential measurement scheme. Unwanted phase fluctuations are diminished by using a stationary reference reflection with respect to the target reflections within the sample[36,38]. Characterization of the reference-subtracted phase noise at a given SNR provides an estimate of the minimum detectable phase change.

To demonstrate the phase stability of the ue-OCT and compare it with a conventional SD-OCT system, a reflective ND filter was chosen as the sample, because each of the two surfaces of the filter provides a reflection peak at two distinctly resolvable depths. For estimating the phase noise of each system, the ND filter sample was imaged by both systems. Each surface of the NDF contributes a reflection to the OCT depth profile, as shown in Fig. 5.a). The unwrapped phase ϕ is extracted from the depth profile. In Fig. 5.b), the difference between the phase of each peak eliminates any oscillatory noise that is common to both surfaces. The NDF is placed at a depth such that both surfaces are in the real plane of the OCT imaging window. The distribution of the unwrapped phase difference Δϕ between the two surfaces is normal. The standard deviation of the phase difference ($\sigma_{\Delta\phi}$), shown in Fig. 5.c), provides an experimental estimate of the phase noise of the system at each SNR.

The experimental phase noise estimates of both a conventional SD-OCT system and the ue-OCT system are compared with the theoretical limit given by Eq. 2 as depicted in Fig.5.d). The SNR at each surface is used to compute a composite SNR. The composite SNR is used to estimate the theoretical limit of the phase noise. The phase noises of the systems at varying signal intensities were characterized by attenuating the sample arm signal with a variable attenuator (absorptive NDF). Fig. 5.d) illustrates the deviation of conventional SD-OCT systems and our ue-OCT system from the theoretical phase noise in a shot-noise limited case at varying SNRs. The conventional SD-OCT system implementation has an average deviation of 0.13 rad with a standard deviation of 82 mrad from the phase noise floor. In our ue-OCT system implementation, we observe an average deviation of 0.12 rad with a standard deviation of 78 mrad from the theoretical noise floor, an insignificant deviation from the conventional system. These experimental results clearly show that our ue-OCT is as phase-stable as a conventional SD-OCT system and therefore, can be used for measuring phase changes, at deeper depths.

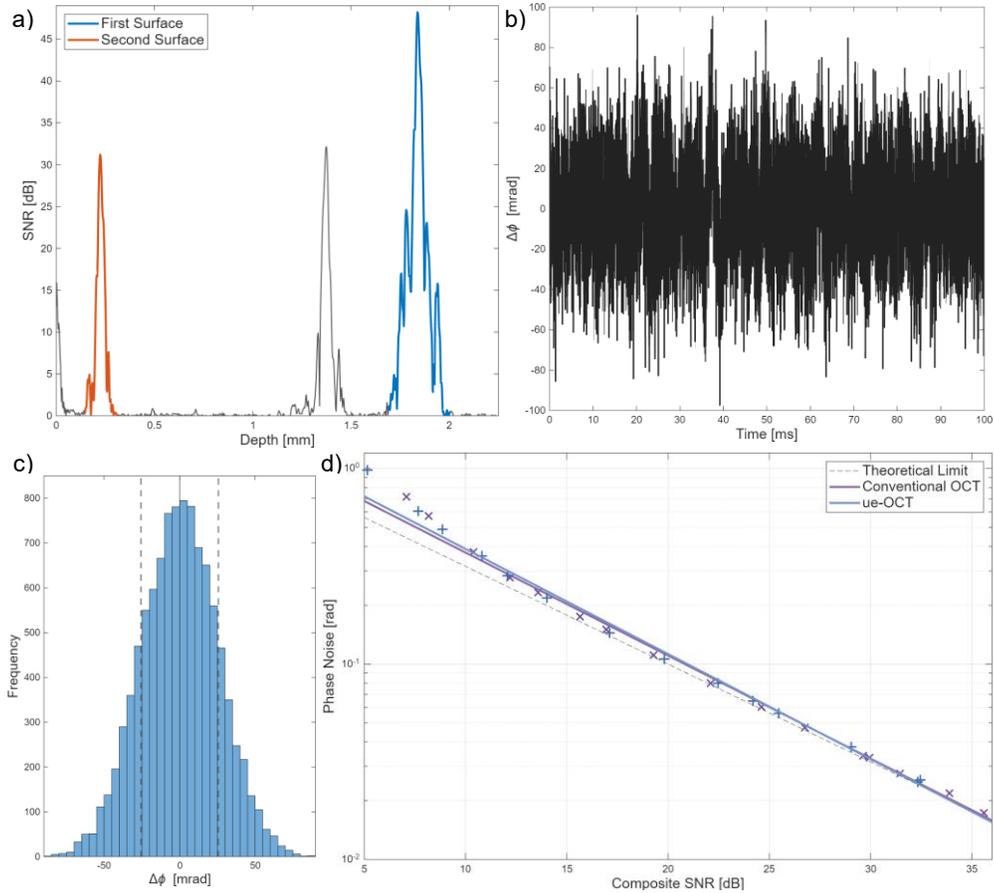

*Figure 5:* a) NDF surfaces in ue-OCT depth reflectivity profile (gray peak is another reflection in the conjugate imaging plane); b) Phase difference between NDF surfaces; c) Phase Noise distribution of phase difference between surfaces; d) Phase noise comparison between our ue-OCT system, a conventional system and the theoretical limit.

## 5. Changes in flow rates can be detected using ue-OCT

Functional imaging using phase-sensitive SD-OCT is realized via relative motion detection between different tissue regions. OCT angiography is commonly used to quantify changes in the relative blood flow rate changes within blood vessels. To demonstrate the ability of the *in situ* ultrasonic lensing configuration to resolve motion, we perfused whole milk through polyethylene tubing. The polyethylene tubing is mounted perpendicular to the optical beam to provide strong reflections from the sample. The tubing has an outer diameter of 1.57 mm and a wall thickness of 265 μm. The tubing is placed flush to the ultrasonic transducer within a water tank thereby acoustically mimicking a blood vessel within tissue. The flow rates that the milk is perfused at are controlled by a syringe pump (InfusionONE; New Era Pump Systems).

To demonstrate that our ue-OCT system can differentiate between different flow velocities relevant in medium and large blood vessels in rodent brain models (1 mm/s – 15 cm/s) [39,40], a syringe pump is used to set the flow rates between 50 μl/min and 15 ml/min. This corresponds to maximum flow velocities between 0.8 mm/s and 24.49 cm/s, based on the inner diameter of the tube (1.14 mm). The selected flow rates also ensure a laminar flow profile, which is parabolic for a circular cross-section[41]. The maximum flow velocities for each flow rate occur at the center of the tube cross-section. The tube containing the flowing fluid is mounted at an angle that is close to perpendicular to the optical beam axis (Fig. 6.a). The reflected signals from the tube interfaces can be effectively detected as the signal intensity is maximized under perpendicular illumination (Fig. 6.b). The interferogram captured by the line scan camera at an A-line period of 94 μs is used to compute the complex depth profile using a k-space Fourier transform. The intensity depth profile of the tube is depicted in Fig. 6.c) for selected flow rates. In addition to the inherent loss in SNR with depth due to the finite size of the line detector pixels, the SNR degrades as a function of depth due to absorption and scattering from the sample. The loss in SNR with depth prevents detection of signals from the entire tube diameter. The changes in SNR due to varying flow velocities are not significant (Fig. 6.b).

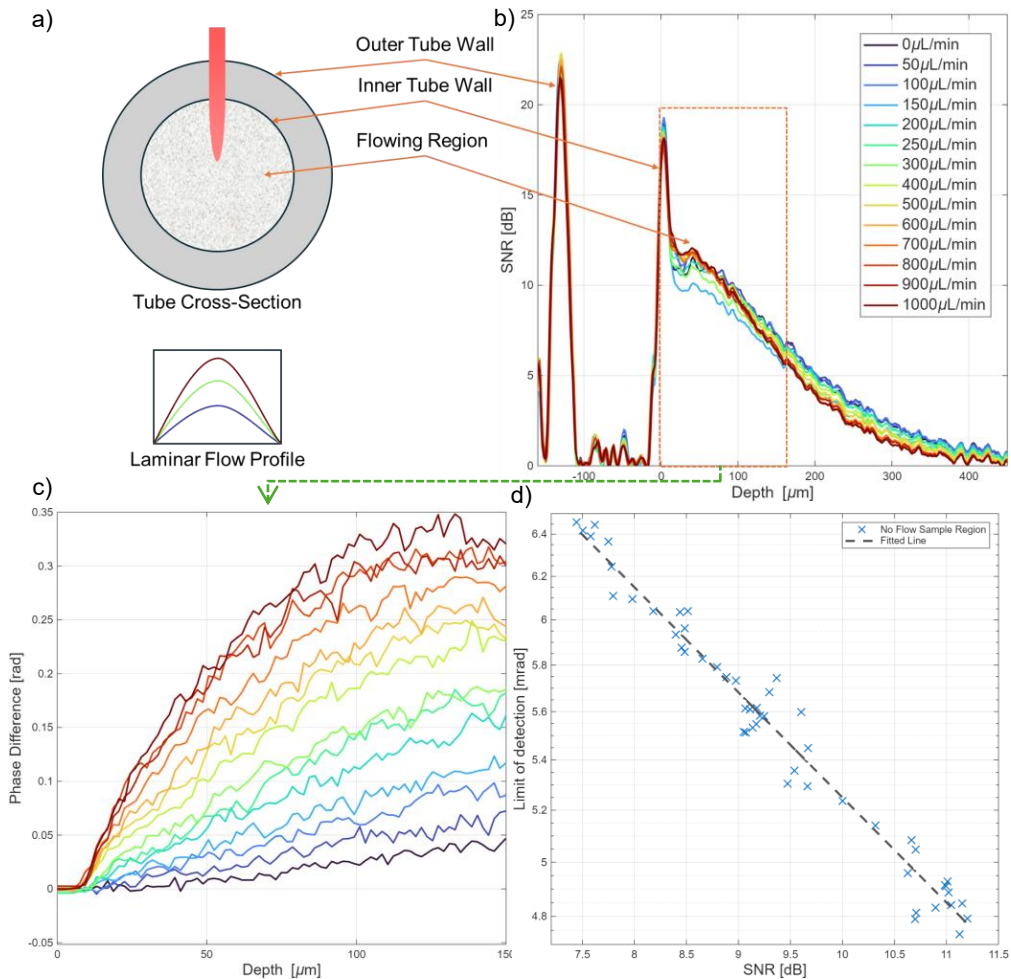

*Figure 6:* a) Sample schematic with laminar flow model; b) Depth intensity profile of tube under different flow conditions; c) Averaged phase difference for flow rates of 50-1000 μL/min; d) Phase noise for no flow condition.

Phase-sensitive SD-OCT leverages the Doppler effect to estimate the velocity of moving particles traveling along the direction of optical beam. The flow velocity of the moving particles is represented as a function of the projection angle between the optical beam and the direction of flow (Doppler angle) as

$$V\cos\alpha = \frac{\lambda_0}{4\pi n \tau}\Delta\phi, \qquad (3)$$

where $\lambda_0$ is the center wavelength of the optical source, $\tau$ is the A-line period of the acquisition camera, and n is the refractive index of the medium. $\phi$ is extracted from the complex depth profile. $\Delta\phi$ is the phase difference between consecutive A-lines at each depth region. From Eq. 3, $\Delta\phi$ is a direct measure of changing velocity. Averaging phase differences ($\langle\Delta\phi\rangle$) can be used to decrease uncertainty in phase difference measurements. The averaged phase differences can characterize flow profiles (Fig. 6.c). The minimum detectable flow rate can be estimated from the standard error of the mean across phase differences ($\sigma_{\langle\Delta\phi\rangle}$) when the sample is not flowing. The limit of detection is inversely proportional to SNR and can be estimated from a fitted line for a range of SNR values before the phase wrapping limit is reached (Fig. 6.d). For example, a flow profile characterized by 20,480 A-line averages has a detection limit of 82 μrad/dB in the selected SNR range. Our ue-OCT can detect flow velocity induced phase changes (Fig. 6.c) following the expected monotonically increasing flow velocity profile described by laminar flow. The corresponding flow velocity detection limit can be computed as a function of SNR. For example, at 10 dB SNR, $\sigma_{\langle\Delta\phi\rangle}$ is 5.25 mrad which represents a flow velocity of 71 $\mu m/s$ in our current system configuration. To detect changing velocities, the outer tube wall identified using the intensity depth profile (Fig. 6.b) is used as a reference phase subtraction to minimize the unwanted bulk phase drift. At regions closer to the center of the tube cross section, the velocities can be more easily differentiated. The ability of ue-OCT to detect changing flow rates confirms the phase stable nature of the system is preserved even under moving sample conditions.

The dynamic range of the system is limited by the fringe washout induced by fast moving samples (Fig. 7). As the flow velocity increases, the modulation frequency of the detected interference fringes increases[42–44]. When the fringe modulation frequency approaches the A-line rate, there is a loss in SNR which contributes to a larger phase noise. This phenomenon is common in conventional SD-OCT and ue-OCT and can be addressed by increasing the A-line period to maximize SNR for faster flow detection. At higher flow velocities, the ue-OCT system with an A-line period of 94 $\mu s$ can detect flow rates with a peak flow velocity of 14.7 cm/s (flow rate of 9 mL/min) before fringe washout reduces SNR beyond where phase stable measurements are possible. While faster exposure times can mitigate fringe washout, the reduction in overall signal intensity within an acquisition must be compensated to detect fast flow in deeper regions. Ultimately, the dynamic range is determined by the phase wrapping limit common to both conventional OCT and our ue-OCT implementation.

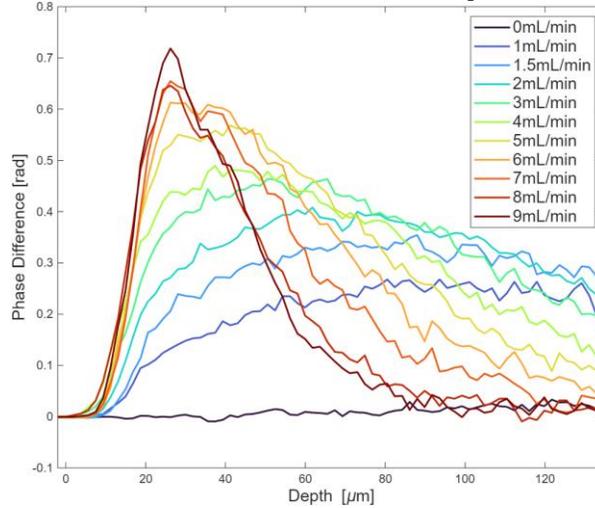

*Figure 7:* Dynamic range of phase sensitive ue-OCT

## 6. Flow detection at depth is enhanced by ue-OCT

The detection of flow in deep tissue using conventional SD-OCT systems is challenging because of SNR loss as a function of depth. We demonstrate improved flow rate detection enabled by ue-OCT using a sample with a flowing medium 3.5 mm deep into a sample. The sample consists of a cover slip mounted on an agar (1%) sample. A polyethylene tube is embedded within agar (Fig. 8.a). The designed sample uses the interface between the glass (cover slip) and agar as a marker for a shallow depth region in the sample. The OCT intensity profile averaged over 1024 A-lines with an A-line period of 995 $\mu s$ shows that at this shallow feature, the SNR is matched for both the conventional SD-OCT and ue-OCT (Fig. 8.b). The SNR matched at the shallow depth region allows for a fair comparison between conventional SD-OCT and ue-OCT for resolving flow at depth. The features in the sample at >3 mm depth are imaged with higher SNR in ue-OCT compared to conventional SD-OCT. As ue-OCT exhibits an extended Rayliegh range, the start of the flow region is imaged with a significant SNR improvement of ~3 dB (Fig. 8.b).

The impact of the higher SNR is significant when detecting flow. The comparison of the average phase differences across 5,115 A-lines between the two systems illustrates the enhancement that ue-OCT can achieve (Fig. 8.c). In the conventional SD-OCT system, in the flow region, the phase noise is significant given the low SNR. For flow rates of 0, 100, and 300 $\mu L/min$, conventional SD-OCT fails to differentiate the flow rates and cannot demonstrate the monotonically increasing trend associated with flow (Fig. 8.c (Left)). Since ue-OCT is as phase stable as conventional SD-OCT and exhibits an extended Rayliegh range, ue-OCT can differentiate flow changes (Fig. 8.c (Right)), that conventional SD-OCT cannot. This demonstrates that ue-OCT relieves the inherent depth-resolution tradeoff of conventional lenses with Gaussian beams used in SD-OCT. The enhanced imaging depth enabled by the *in situ* optical waveguiding coupled with phase stability enables distinguishing phase changes and flow rate changes at deeper tissue regions, otherwise undistinguishable by conventional OCT.

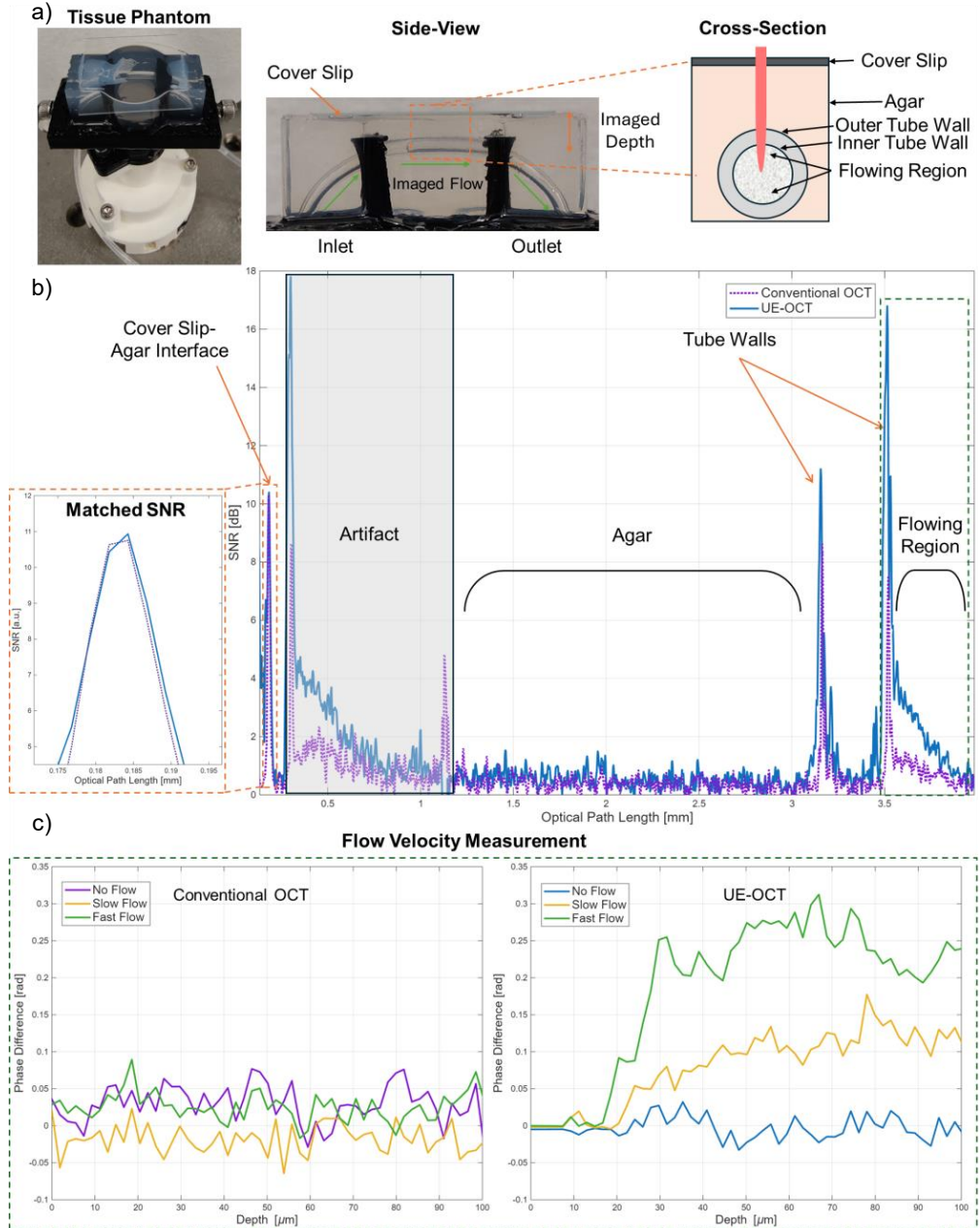

*Figure 8:* a) Sample for flow at depth system comparison; b) Depth intensity profile comparison between conventional OCT and ue-OCT (call out: matched SNR at shallow depth); c) ue-OCT can detect flow velocity changes at depth otherwise not possible with a matched conventional OCT implementation.

## 7. Discussion and Conclusion

Our *in situ* ultrasonically sculpted optical waveguiding technique offers great potential in applications that would benefit from *in situ* light guiding within tissue samples. The ability of the ultrasonically sculpted waveguides to spread or confine a focal volume provides significant advantages over the geometrical limitations imposed by external lenses. Most importantly, this technology when combined with SD-OCT can be used for functional imaging via phase sensitive detection. The ultrasonic transducer used in our implementation is a cylinder radiating along its radial mode defined by a zero-order Bessel function of the first kind. The phase of the standing wave that interacts with the optical source can be selected by synchronizing the optical source and electrical signal supplied to the ultrasound transducer. The temporal dynamics of the pulsed optical source and the dynamic ultrasonically sculpted refractive index profile determines the effective spread of optical intensity throughout the sample during one acquisition. In our implementation, we shutter the optical source at a 6% duty cycle and synchronize our optical source with the peak of the ultrasonic standing wave for peak SNR. The current implementation of our *in situ* ultrasonic lensing system using a pulsed laser source (6% duty cycle) effectively uses a maximum optical power of 0.6 mW incident on the sample. While the input power projected onto the sample is rather low, which limits the sensitivity of the system, a more powerful light source can be used to improve the effective optical power and sensitivity.

The development of *in situ* ultrasonically sculpted optical waveguides can also be extended to perform volumetric imaging with ue-OCT. Through multi-element ultrasonic transducers, *in situ* lateral beam steering can be implemented in ue-OCT for three-dimensional functional imaging [45,46]. Integrating electronically controlled beam steering would also allow for increased scan speeds over longer settling time galvo mirrors used in conventional OCT. The refractive index gradient created by the *in situ* ultrasonically sculpted optical waveguides can be tuned by changing voltage applied across the ultrasonic transducer elements. The tunable nature allows for rapid and inexpensive prototyping of ue-OCT systems. These extensions of ue-OCT could provide further improvements on conventional SD-OCT systems for structural and functional imaging.

The improvement in SNR over depth enabled by using our ultrasonically sculpted optical waveguides help enhance multiple biological imaging modalities in OCT. Inherently, structural imaging of biological tissue benefits from increased SNR as a function of depth. The extension in depth enables probing of deeper tissue regions with improved contrast. Additionally, the higher SNR retrieved from each depth within the tissue allows for a lower phase noise at the same depth compared to conventional phase-sensitive SD-OCT systems. We have demonstrated that this increased phase stability at deeper depths results in detection of flow using ue-OCT, otherwise not possible using conventional SD-OCT. These experimental results imply that in biological samples where attenuation due to scattering and absorption of tissue is not dominant, flow at deeper tissue regions can be resolved with a higher sensitivity. Improved sensitivity would allow for improved monitoring of disease progression as well as efficacy of therapeutics. The major implication of flow detection at deeper regions using ue-OCT could lead to detecting disease onset, previously not possible.

## Acknowledgements

This research was supported in part by the NSF award 2526580, and Phillip and Marsha Dowd Graduate Fellowship for Lloyd Lobo.

## Disclosures

The authors declare that there are no conflicts of interest

## Data availability

The data underlying the results presented in this paper are not publicly available at this time but may be obtained from the authors upon reasonable request.